# Computational Sociology of Humans and Machines; Conflict and Collaboration

Taha Yasseri [1,2,3]

[1] School of Social Sciences and Philosophy, Trinity College Dublin, Dublin, Ireland
[2] Faculty of Arts and Humanities, Technological University Dublin, Dublin, Ireland
[3] School of Mathematics and Statistics, University College Dublin, Dublin, Ireland

taha.yasseri@tcd.ie

**Abstract**

This Chapter examines the dynamics of conflict and collaboration in human-machine systems, with a particular focus on large-scale, internet-based collaborative platforms. While these platforms represent successful examples of collective knowledge production, they are also sites of significant conflict, as diverse participants with differing intentions and perspectives interact. The analysis identifies recurring patterns of interaction, including serial attacks, reciprocal revenge, and third-party interventions. These microstructures reveal the role of experience, cultural differences, and topic sensitivity in shaping human-human, human-machine, and machine-machine interactions. The chapter further investigates the role of algorithmic agents and bots, highlighting their dual nature: they enhance collaboration by automating tasks but can also contribute to persistent conflicts with both humans and other machines. We conclude with policy recommendations that emphasize transparency, balance, cultural sensitivity, and governance to maximize the benefits of human-machine synergy while minimizing potential detriments.

# Introduction

This Chapter explores the dynamics of conflict and collaboration among humans and machines. It begins with a question posed in Morris Zelditch Jr.'s 1969 article, *Can You Really Study an Army in the Laboratory?* Zelditch (1969) argues that while laboratory experiments can provide valuable insights, they may not fully capture the nuances and complexities of real-world organizational dynamics. He highlights the inherent challenges of studying intricate social phenomena, such as conflict and collaboration, within controlled scientific settings. Laboratory studies, although effective in offering a framework for isolating and analyzing variables, often struggle to account for the contextual richness and emergent properties of interactions in large-scale, real-world systems. Drawing parallels to other disciplines, Zelditch emphasizes the need for methodological innovations that address this tension between experimental control and ecological validity. This chapter builds on these ideas, extending them to the study of human-human human-machine and machine-machine interactions, where similar challenges arise in capturing the intricate interplay of cooperation and competition in increasingly complex socio-technical systems.

Although significant changes have occurred since 1969, research in this area remains sparse, particularly at the micro-level. Existing studies have predominantly focused on collaboration, largely because positive interactions, such as friendship, mentorship, and teamwork, are easier to observe and analyze, and micro-level studies of social networks have therefore yielded significant insights into positive interactions among humans in social groups (Freeman, 2004). In contrast, research on negative interactions—such as disagreement, disapproval, or distrust—remains underdeveloped. Most analyses of conflict are restricted to macro-level phenomena, such as national, regional, or international conflicts. At the level of one-to-one interactions, data remain extremely limited. For instance, unlike platforms such as Facebook or X that explicitly document positive interactions in the shape of friendship, followership, and *likes*, no equivalent social network exists where users can document adversarial relationships or declare their 'enemies.' This absence has hindered the study of negative interpersonal interactions at the micro level. When it comes to machines, the research is even more sparse.

To address this limitation, we turned to Wikipedia, the largest collaboratively written encyclopedia in human history. Wikipedia contains more than 60 million articles across more than 300 language editions and serves as an extensive case for studying collaborative and adversarial interactions. While Wikipedia is widely regarded as a successful example of large-scale, internet-based collaboration, its editorial process is far from harmonious. Wikipedia's decentralized editorial team consists of millions of contributors with diverse opinions, intentions, motivations, and knowledge. These contributors often rely on conflicting or contradictory sources, which creates fertile ground for editorial disputes.

Disagreements often occur when editors cannot agree on what content should be included in an article, whether a topic merits its own article, how an article should be titled or structured, or other editorial decisions. Such conflicts lead to "edit wars," wherein editors repeatedly override each other's contributions (Sumi et al., 2011a;Sumi et al., 2011b). These disputes can pertain to contentious current events, where reliable information is still emerging, or historical topics, which may seem well-documented yet remain subject to competing interpretations. Editors' personal biases, intentions, and selective use of sources often exacerbate these disagreements.

## Collaboration and conflict among humans

Analyzing editorial conflicts is inherently complex. Anyone who has collaboratively written a text with colleagues knows how difficult it becomes to trace contributions, deletions, and the evolution of the document. To simplify this problem in Wikipedia, we focused on "reverts." Reverts are a specific type of edit enabled by Wikipedia's software that allows users to instantly restore a previous version of an article. Originally designed as a mechanism to combat vandalism—where editors could quickly undo disruptive changes—the revert function is frequently employed in disputes between contributors, even when all involved parties are experienced editors.

Reverts provide a straightforward way to identify negative interactions between pairs of editors. Computationally, they are easy to track: by comparing different versions of an article, one can detect reverts whenever two versions are identical, indicating that all edits in between have been undone. This method allows us to construct networks of binary, negative interactions among editors involved in reverts.

### The big picture

We analyzed Wikipedia's first ten years (2001-2011) across 13 different language editions, identifying approximately 4.7 million reverts. Figure 1 illustrates an example of these interactions, which exhibit characteristics similar to social networks, such as a skewed degree distribution. However, this network represents negative interactions, which diverge from typical social network dynamics. For instance, the balance theory predicts that "the enemy of my enemy is my friend" (Cartwright & Harary,1956). Therefore, we do not expect to observe many triangular structures in a network of reverts, unlike typical social networks.

The first key question we explored was: which topics generate the most controversy across different Wikipedia language editions? Table 1 presents the most contested articles in various languages. Some articles are predictably controversial, such as those related to politics, religion, and other socially sensitive issues—subjects that one might avoid in polite conversation. However, unexpected patterns also emerge. For example, in the Spanish-language Wikipedia, many contested articles pertain to football clubs. However, it is worth noting that some of these football clubs carry strong political associations, which may contribute to these conflicts. Similarly, in the Czech-language Wikipedia, we observe heightened controversy around topics related to sex and sexuality. This pattern reflects the cultural and societal transformations experienced by the Czech Republic and other former Eastern Bloc countries over recent decades. These findings offer unique insights into the cultural priorities, sensitivities, and obsessions of different language communities that contribute to Wikipedia. By examining the most contested topics and their underlying dynamics, we gain a deeper understanding of how conflict and collaboration manifest in this global, internet-based platform.

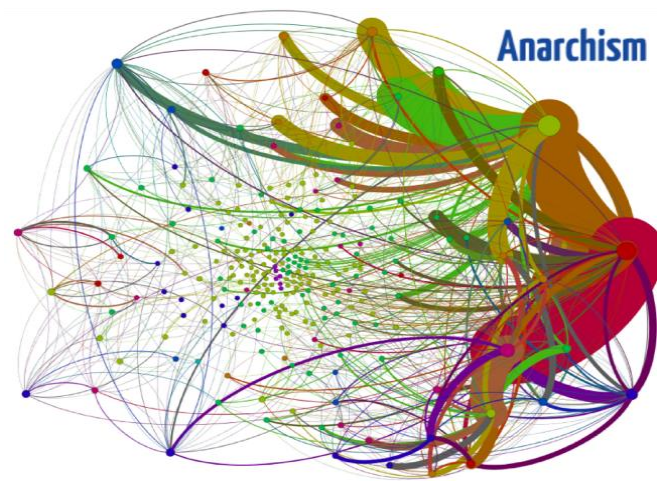

**Figure 1.** A network representation of reverts in the history of the article on "Anarchism" in English WP. Nodes are editors, and links represent reverts. The size of the nodes is proportional to the total number of reverts in which the editor is involved, and the width of the links is proportional to the number of total reverts between their corresponding pair of editors. Few nodes are strongly connected, whereas most of the nodes are connected only through weak links. *Completed triangles* are clearly under-represented. *The graph is generated by Gephi*. Source: Yasseri & Kertész (2013).

## Dynamics of conflict and collaboration

Now, we examine the dynamics of controversy over time. Different articles exhibit distinct patterns of conflict. Some articles begin peacefully, experience a tense period of editorial disagreement, and eventually reach a consensus on their content, e.g., the *Bombing of Dresden*. Other articles, however, such as *Anarchism*, experience no periods of peace; their editorial histories are essentially ongoing wars. A third group of articles, like the example of *Japan*, demonstrates oscillating periods of peace and conflict that occur sequentially. This dynamic often emerges because these articles cover evolving topics, such as countries, where new events and information can trigger renewed editorial activity. Editors may reach a temporary consensus, followed by renewed conflict as new developments arise, leading to a burst of reverts (Yasseri et al., 2012).

**Table 1** Top-10 most controversial articles in different language editions of Wikipedia. Titles in *italic* are literally translated; the rest are the titles of the sister articles in English Wikipedia. en: English, de: German, fr: French, es: Spanish, cs: Czech, hu: Hungarian, ro: Romanian, ar: Arabic, fa: Persian, and he: Hebrew. Source: Yasseri et al., (2014).

| en | de | fr | es | cs |
|---|---|---|---|---|
| Anarchism | Scientology | Unidentified flying object | Club América | Psychotronics |
| Muhammad | 9/11 conspiracy theories | Jehovah's Witnesses | Opus Dei | Telepathy |
| LWWEe | *Fraternities* | Jesus | Athletic Bilbao | Communism |
| Global warming | Homeopathy | Sigmund Freud | Andrés Manuel López Obrador | Homophobia |
| Circumcision | Adolf Hitler | September 11 attacks | Newell's Old Boys | Jesus |
| United States | Jesus | Muhammad al-Durah incident | FC Barcelona | Moravia |
| Jesus | Hugo Chávez | Islamophobia | Homeopathy | Sexual orientation change |
| Race and intelligence | Minimum wage | God in Christianity | Augusto Pinochet | Ross Hedvíček |
| Christianity | Rudolf Steiner | Nuclear power debate | Alianza Lima | Israel |
| **hu** | **ro** | **ar** | **fa** | **he** |
| *Gypsy Crime* | FC Universitatea Craiova | Ash'ari | Báb | Chabad |
| Atheism | Mircea Badea | *Ali bin Talal al Jahani* | Fatimah | Chabad messianism |
| Hungarian radical right | Disney Channel (Romania) | Muhammad | Mahmoud Ahmadinejad | 2006 Lebanon War |
| Viktor Orbán | Legionnaires' rebellion & Bucharest pogrom | Ali | People's Mujahedin of Iran | B'Tselem |
| Hungarian Guard Movement | Lugoj | Egypt | Criticism of the Quran | Benjamin Netanyahu |
| Ferenc Gyurcsány's speech in May 2006 | Vladimir Tismăneanu | Syria | Tabriz | Jewish settlement in Hebron |
| *The Mortimer case Hungarian Far-right* | Craiova | Sunni Islam | Ali Khamenei | Daphni Leef |
| Jobbik | Traian Băsescu | Yasser Al-Habib | Massoud Rajavi | Beitar Jerusalem F.C. |
| Polgár Tamás | Romanian Orthodox Church | Arab people | Muhammad | Ariel Sharon |

## Modeling Conflict and Collaboration

Next, we model the behavior of Wikipedia editors using the agent-based modeling paradigm. For this, we adopted the *Bounded Confidence* model, originally introduced by Deffuant et al. (2000). In the basic version of this model, there are *N* agents, each with an opinion represented as a scalar value between 0 and 1. The interaction rule is as follows: when two agents meet if the difference between their opinions is smaller than a predefined threshold or tolerance level ($\varepsilon$), they adjust their opinions toward each other. While this process does not have to be symmetric, for simplicity, we assume that both agents update their opinions toward each other by the same amount.

The outcomes of the basic model depend on the value of $\varepsilon$. If $\varepsilon$ is sufficiently large (i.e., agents have high tolerance for differing opinions), the model converges to a state of complete consensus, where all agents share the same opinion. Given the symmetry of the model, this opinion converges to the midpoint value of 0.5. Conversely, when $\varepsilon$ is small (i.e., agents have low tolerance), opinions fragment into multiple groups. Within each group, agents achieve consensus, but the groups remain isolated, with no possibility of reconciling their differences. This fragmentation is stable because the opinion distances between groups exceed the tolerance level $\varepsilon$. This scenario corresponds to a state of permanent conflict across groups—a "never-ending war."

However, the basic Bounded Confidence model does not fully capture the behavior of Wikipedia editors because it overlooks a critical element: the article itself. The article can be conceptualized as a separate agent, with its own "opinion" represented as a value between 0 and 1. The interaction rule between editors (agents) and the article is defined as follows: when an agent interacts with the article, they compare their opinion to the opinion reflected in the article. If the difference exceeds their tolerance level ($\varepsilon_A$), the agent edits the article to bring it closer to their own opinion. If the article's opinion is already within the agent's tolerance, the agent does not edit the article but instead adjusts their own opinion toward the article's content.

We ran simulations of this extended model, beginning with the standard Bounded Confidence framework until multiple opinion groups formed. At this point, we introduced the article as an agent and allowed editors to interact with it. The inclusion of the article enables long-range, indirect interactions among editors who would otherwise no longer engage with one another. Surprisingly, these indirect interactions eventually drive all editors and the article toward consensus (Török et al., 2013).

It is important to note that the resulting consensus does not necessarily reflect the average of all initial opinions. Introducing the article breaks the symmetry of the original model, which has significant consequences: even when consensus is achieved, the article's content may not accurately represent the full range of opinions among editors, let alone the objective truth (Iñiguez et al., 2014).

While this model explains the consensus-building process, it does not account for the "never-ending wars." The model lacks an essential feature: the renewal of agents. In Wikipedia, the pool of editors contributing to an article is not static; new editors arrive with fresh perspectives, and old editors leave. To address this, we introduced agent renewal into the model. At each step, new agents with randomly assigned opinions (between 0 and 1) are added, while existing agents are removed at a certain rate $P$. By fine-tuning the parameters $P$ (renewal rate), $\varepsilon$ (tolerance between agents), and $\varepsilon_A$ (tolerance for the article), we successfully replicated our empirical observations through agent-based modeling (Török et al., 2013).

The purpose of this simulation extends beyond replicating observed behavior. It also allows us to systematically explore the parameter space and identify policies to manage editorial conflict. For instance, Figure 2 demonstrates the effects of varying $\varepsilon_A$ (tolerance for article content) and the size of the editorial pool $N$ and renewal rate of the editor pool $P$. Different parameter values produce distinct phases: articles may remain in a peaceful state, enter a period of conflict due to external events increasing N, or become permanently contentious when $\varepsilon_A$ decreases due to increase in topic's sensitivity.

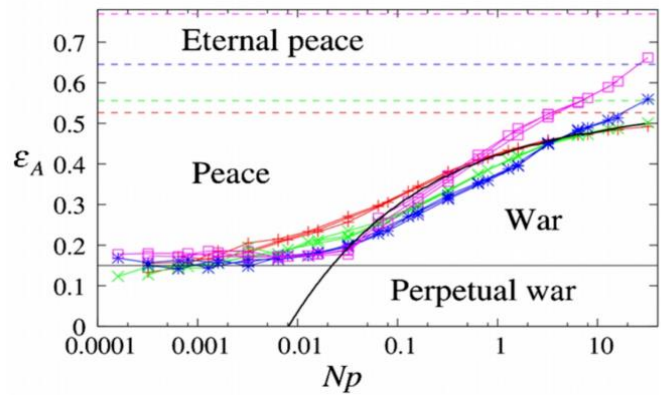

**Figure 2.** The phase diagram in the case of agent replacement. Different phases are separated by simulation results and numerical analysis of the model. Adapted form (Török et al., 2013).

These findings can inform editorial policies. For example, freezing an article—effectively reducing the renewal rate $P$—can stabilize its content and minimize conflict. While such policies may appear contrary to the principle of Wikipedia as a "free encyclopedia that anyone can edit," they can ensure stability in highly contested articles. This approach aligns with policies already adopted by Wikipedia editors.

Having established a working model, we explored additional policies. For example, we examined the effects of banning editors with extreme, inflexible opinions. Counterintuitively, these editors often expedite consensus (Rudas et al., 2017). Their extreme positions act as a catalyst, disrupting the system sufficiently to allow it to settle into equilibrium more quickly, even if this resulting consensus does not reflect the truth or even results in a democratic compromise.

## Status, power, and interaction motifs

Finally, we examine the micro-dynamics of interactions among Wikipedia editors. Figure 3 presents all the reverts that occurred within a single day of Wikipedia history, specifically January 15, 2010. In this figure, reverts between two editors are not collapsed into a single edge but are shown distinctively. Upon analyzing this network, it quickly becomes apparent that it is not random; instead, we observe recurring microstructures and patterns, which we refer to as *motifs*. When the network is colored based on these motifs, we find that certain motifs cluster together, appearing closer to one another, while others are more spatially distant. These observations suggest the presence of underlying forces that regulate and shape these patterns.

To study these motifs, we focused on three key properties: their *prevalence* (how frequently they occur), their *pace* (the temporal order and speed with which elements of the motif follow one another), and their *structure* (whether the attributes of the editors, such as experience, explain the formation of these motifs).

One of the most prevalent motifs we identified is the *serial attack* motif. This occurs when one editor repeatedly reverts another editor's contributions. Analysis of this motif revealed that it often involves an asymmetry in editor experience; specifically, a more experienced editor, exhibiting greater confidence, repeatedly targets a less experienced editor. Additionally, these serial attacks demonstrate a high pace, meaning that the reverts occur much more rapidly than expected under a null model where reverts are assumed to be uncorrelated over time.

Another frequently observed motif is the *ABBA* motif, which we refer to as *revenge*. In this pattern, an attack (revert) by one editor is reciprocated by the other editor. Unlike the serial attack motif, the ABBA motif tends to occur between editors with similar levels of experience, where neither editor holds seniority over the other. This equality in experience appears to encourage reciprocity, as neither editor concedes to the other. As with the serial attack motif, the ABBA motif also exhibits a high temporal pace, indicating that reverts occur more quickly than expected under a null model. We observed fewer instances of the ABBA motif in the Chinese and Japanese Wikipedia editions. This finding may reflect cultural differences in the editing behavior (Yasseri, Sumi, Kertész, 2012; Yasseri, Quattrone & Mashhadi, 2013) and conflict resolution strategies of Wikipedia editors in these language communities.

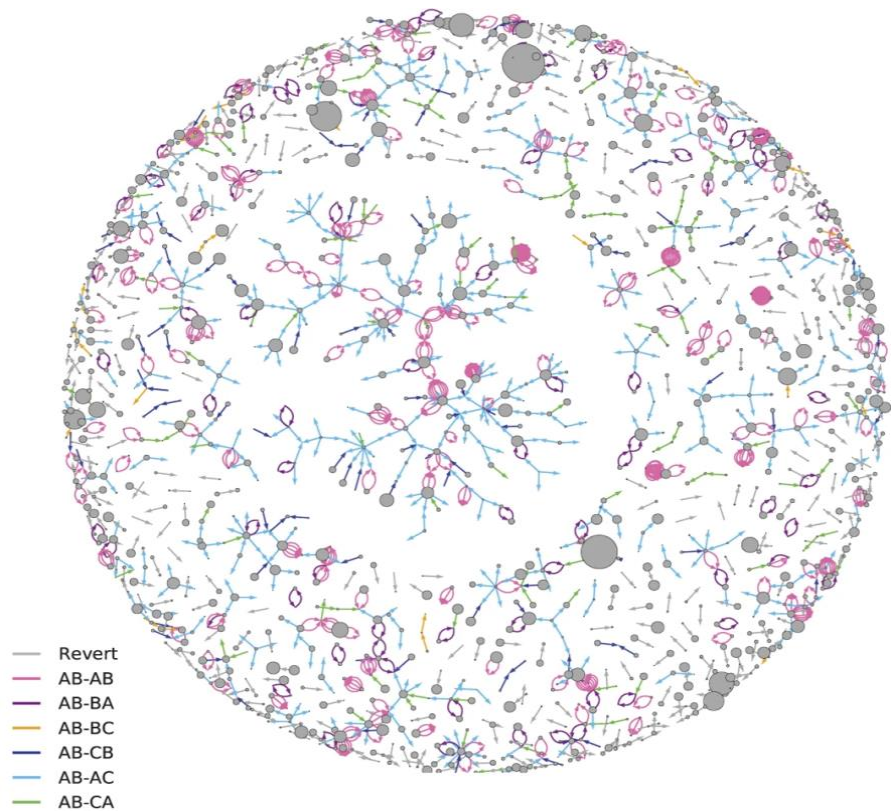

**Figure 3.** An example network of all reverts was done on Wikipedia within one day (January 15, 2010). The six temporal motifs that we investigate are color-coded. A link points from the reverter to the reverted user. The area of the nodes corresponds to the total number of edits by the user. Source: Tsvetkova, García-Gavilanes & Yasseri (2016).

The third motif to mention is the *third-party defense* motif. In this pattern, Editor A reverts Editor B, but a third editor, Editor C—who is typically much more experienced than both Editor A and Editor B—intervenes to revert Editor A, effectively defending Editor B. The temporal fingerprint of this motif reveals a significant temporal correlation, indicating that the sequence of reverts does not occur randomly but reflects a deliberate and structured response.

To summarize this section:

1. Collaboration facilitates consensus: In our agent-based models, the introduction of a *common product* (such as an article) leads editors who would otherwise disengage from consensus-building to converge toward similar opinions. The shared goal of editing the article fosters interactions that promote agreement.

2. Banning extreme editors delays consensus: In this context, "extreme editors" refer to those whose opinions are significantly closer to 0 or 1 compared to the majority. Although banning such editors may initially seem beneficial, our findings indicate that it only postpones the process of consensus building.

3. Editor experience influences microstructures: The status and experience of editors strongly explain the prevalence and structure of observed motifs. For example, serial attacks are often initiated by experienced editors, while revenge motifs occur between editors of similar experience levels.

With these insights into the micro-dynamics of editorial interactions, we now turn our attention to the role of bots in Wikipedia editing.

## Collaboration and conflict among machines

Internet bots, also referred to as internet robots, algorithms, AI agents, or AI algorithms, all represent the same fundamental concept: algorithmically driven agents that interact with humans, other bots, and the information on web-based platforms. Here, we focus specifically on bots that edit Wikipedia.

While the prevalence of bots, particularly those powered by large language models and tools like ChatGPT, has increased dramatically in recent years, automated or semi-automated tools have been employed on Wikipedia since as early as 2000. These algorithmically driven tools have been used by Wikipedia editors to enhance the quality and content of articles.

Wikipedia bots serve a variety of functions. Basic bots are used to undo vandalism, enforce editorial bans, correct spelling errors, and create inter-language links. More sophisticated bots automate content imports, while advanced algorithms generate content and ensure the quality of human contributions. Although much of the editorial activity performed by bots is not immediately visible to casual observers, in certain Wikipedia language editions, bots are responsible for up to 50% of all edits (Tsvetkova et al., 2017).

The primary difference between Wikipedia bots and bots on social media platforms lies in their purpose. Wikipedia bots share a common overarching goal: to improve the quality of articles. Furthermore, many bots rely on shared underlying code and technology. In this sense, one could argue that Wikipedia bots are simply different manifestations of the same algorithmic entity. However, despite their shared technological foundation, the behavior of bots can vary significantly depending on the users running them and the Wikipedia environment in which they operate.

### Dynamics of bot-bot interactions

Bots, like human editors, are involved in reverts—they not only revert the contributions of humans and other bots but get also reverted themselves. This is particularly true for bots that are still in their trial phase and lack experience. In some language editions, up to 8% of bot edits are reverted by humans, while approximately 2% are reverted by other bots. This observation highlights the complexity of bot interactions within Wikipedia. Despite being deterministic and rule-based, bots are still subject to reversions due to conflicting guidelines and different operating instructions.

To better understand bot interactions, we analyzed the sequence and dynamics of reverts between pairs of bots and compared them to reverts between human editors. The key differences we observed are as follows:

1. Pace of Reverts: Bot-bot reverts occur at a much slower pace than human-human reverts. Unlike humans, bots do not act based on emotions, preferences, or personal motivations. Instead, reverts between bots are primarily driven by differing guidelines and instructions. Because bots traverse Wikipedia content in a near-random manner, the probability of two bots with conflicting guidelines encountering each other dictates the frequency of bot-bot reverts. This results in slower, less temporally clustered interactions compared to the rapid bursts of reverts often observed among human editors' conflicts.

2. Persistence Over Time: Bot-bot reverts are significantly more persistent over time compared to human-human reverts. Human editors are prone to "forget" or "forgive" over time, and their priorities often shift, leading to a natural cooling-off period after heated conflicts. In contrast, bot-bot interactions can persist indefinitely, as bots strictly adhere to their programming. Remarkably, we observed pairs of bots reverting each other for years without detection and intervention by human editors.

3. Absence of Power Dynamics: In human interactions, status, and experience play a critical role in shaping editorial conflicts, as discussed in the previous section. However, these power dynamics are largely absent in bot-bot interactions. Bots do not account for the experience or seniority of other bots; they simply execute their programmed instructions. If a bot identifies content that it deems incorrect, it will revert it, regardless of the origin or experience level of the other bot.

## Cultural and linguistic differences

The cultural differences observed among human editors in different Wikipedia language editions are also reflected in bot behavior (Tsvetkova et al., 2017a). This observation is particularly surprising, as technology itself is typically expected to operate independently of cultural factors. However, differences in editorial guidelines, regulatory practices, linguistic nuances, and the habits of human editors who program and operate bots contribute to the observed variation in bot interactions across different languages.

For example, we found that German Wikipedia exhibits the lowest rate of bot-bot reverts, whereas Portuguese Wikipedia shows a bot-bot revert ratio seven times higher than that of the German edition. These findings demonstrate that the same underlying technology can yield vastly different outcomes depending on the environment in which it operates and the humans responsible for its implementation.

In summary, Wikipedia bots play a crucial role in improving the platform's content but are also subject to complex dynamics of conflict and collaboration. Despite their deterministic nature, bots interact in ways that reflect human influence, cultural context, and conflicting operational guidelines; their collective emergent behavior differs from the simple accumulation of individual bots' outputs, a defining characteristic of complex systems (Bianconi et al., 2023).

## Humans and machines in a team

The integration of bots into collaborative platforms like Wikipedia has redefined the dynamics of human-human interaction. While bots are primarily designed to assist with tasks such as vandalism detection, content moderation, and quality assurance, their deployment can influence how humans interact with each other in complex and unexpected ways. This section examines the implications of bot deployment for human teamwork, focusing on collaboration, competition, and the emergent patterns of human-machine synergy (Tsvetkova et al., 2017b).

### Bots as Facilitators of Collaboration

Bots play a pivotal role in reducing the cognitive and operational load on human editors by automating repetitive tasks. By addressing mundane yet essential activities—such as reverting vandalism, correcting spelling errors, and enforcing editorial guidelines—bots allow human editors to focus on higher-level contributions. This division of labor can foster a more efficient and productive collaborative environment (Tsvetkova et al., 2024). Furthermore, AI-driven bots can act as "co-participants," enhancing collective intelligence by providing insights, content suggestions, and decision-making support (Yasseri & Menczer, 2023).

Research on hybrid human-AI systems demonstrates that machines can complement human capabilities when appropriately integrated into teams. For instance, bots can process vast amounts of data more quickly and accurately than humans, offering valuable input that humans may use to make better-informed decisions. Cui and Yasseri (2024) propose a multilayer framework of human-AI collaboration, where bots operate alongside humans in cognitive, informational, and physical layers, enabling teams to achieve outcomes that neither could attain alone.

## Impact on Human-Human Interaction

While bots facilitate collaboration, their presence can also introduce subtle changes in human-human dynamics. A notable effect is the potential reduction in direct human-to-human communication, which has been shown to be vital for effective collaboration (Straub, Tsvetkova& Yasseri, 2023). As bots mediate editorial conflicts by automating reverts and enforcing guidelines, they may inadvertently discourage human dialogue, which is essential for consensus-building and collaborative problem-solving (Cui & Yasseri, 2024). Other work has shown the collaboration among human pairs is diluted once one of the editors starts deploying a bot (Vedres et al., 2025).

Bots can also reshape group hierarchies. Experienced human editors often rely on their social status to influence discussions and resolve disputes. However, bots, which operate independently of human biases and hierarchical norms, can disrupt these dynamics. For instance, bots are equally likely to revert edits made by experienced or novice editors, diminishing the traditional role of seniority in human interactions (Yasseri et al., 2012). This phenomenon can lead to a perception of fairness, but it may also generate frustration among editors accustomed to hierarchical structures.

## Trust and Cooperation in Human-Machine Teams

The perception of bots by human editors plays a crucial role in determining the effectiveness of human-machine collaboration. While bots are generally seen as neutral and unbiased, their involvement in editorial processes can lead to mixed outcomes. Research finds that humans are more willing to exploit AI systems than cooperate with them, particularly when bots are perceived as tools rather than collaborative partners (Karpus et al., 2021); Bazazi, Karpus & Yasseri, 2024). This tendency stems from the lack of emotional reciprocity and moral accountability in human-machine interactions.

Nevertheless, bots can also foster trust and cooperation under certain conditions. Research shows that humans respond positively to bots that display human-like traits, such as transparency and predictability (Kiesler, Sproull & Miller, 1996). By designing bots that are more communicative and adaptive to human workflows, it is possible to mitigate resistance and enhance collaborative outcomes.

# Policy Implications and the Path Forward

The integration of bots into collaborative platforms like Wikipedia presents both opportunities and challenges. It is essential to strike a balance between automation and human oversight to maximize the benefit. Key policy recommendations include:

- Transparent Design: Ensuring that bots operate transparently, with clear guidelines and visible decision processes, can build trust and reduce resistance among human editors (Burton et al., 2024).
- Human-AI Synergy: Bots should be designed as tools to augment, rather than replace, human contributors. Collaborative interfaces that promote meaningful interactions between humans and bots can enhance collective intelligence (Cui & Yasseri, 2024).
- Monitoring and Governance: Establishing robust review mechanisms to monitor machine activity can minimize errors and prevent synergistic cycles of conflict (Yasseri, 2025).
- Cultural Adaptation: Recognizing and accounting for cultural and contextual differences can improve the effectiveness of bots across diverse Wikipedia language editions.

## Conclusion

This chapter explored the dynamics of conflict and collaboration among humans and machines within the framework of Wikipedia, the largest collaborative knowledge platform in history. By analyzing human interactions, machine involvement, and their combined influence, we unveiled unique insights into the micro- and macro-level patterns of cooperation and conflict.

The interplay between human editors and machine agents on Wikipedia provides a compelling lens through which to study conflict, collaboration, and emergent patterns of interaction in the digital age. As bots and AI systems become increasingly integral to collaborative environments, understanding their influence on human dynamics is essential. By fostering thoughtful policies, embracing transparency, and promoting synergy between humans and machines, we can harness the full potential of these partnerships to improve knowledge production while preserving the collaborative spirit that defines platforms like Wikipedia.

In conclusion, collaboration, even in contentious and polarized environments, remains a cornerstone of productive human-machine systems (Eide et al., 2016). By leveraging the strengths of both human and algorithmic agents while addressing their limitations, we can design systems that promote consensus, reduce conflict, and enhance collective outcomes. These lessons are critical not only for improving collaborative knowledge platforms but also for addressing broader societal challenges related to misinformation, polarization, and digital cooperation.

## Further Reading

We refer the reader to Yasseri Kertész (2013) for a detailed examination of value production in collaborative environments, using Wikipedia as a case study to explore sociophysical dynamics and online cooperation. Yasseri & Menczer (2023) provide a forward-looking discussion on how collaborative designs and community-based moderation could address the challenges of social media, fostering healthier marketplaces of ideas. For insights into the evolving role of artificial intelligence in collective decision-making, Cui & Yasseri (2024) offer an analysis of AI-enhanced collective intelligence and its potential applications. Tsvetkova et al. (2024) propose a new sociological framework that examines the interplay between humans and machines, highlighting the shifting boundaries of social behavior in the age of AI. Lastly, Burton et al. (024) examine how LLMs influence decision-making and collaboration within collective intelligence systems, including their application in Wikipedia.

## Acknowledgment

I would like to thank my co-authors of the core papers reviewed in the chapter: Robert Sumi, András Rung, András Kornai, János Török, János Kertész, Gerardo Iñiguez, Kimmo Kaski, Maxi San Miguel, Anselm Spoerri, Mark Graham, Milena Tsvetkova, Ruth Garcia, Csilla Rudas, Olivér Surányi, Filippo Menczer, Luciano Floridi and Hao Cui. The research conducted in this publication was funded by the Irish Research Council under grant number IRCLA/2022/3217, ANNETTE (Artificial Intelligence Enhanced Collective Intelligence).